\documentclass[aps,prl,preprint,groupedaddress]{revtex4}
\usepackage{graphicx}

\begin{document}
\title{ Thiol density dependent classical potential for methyl-thiol on a Au(111) surface}
\author{Byoungseon Jeon}
\affiliation{Department of Applied Science, University of California, Davis, 
CA 95616 \\ Theoretical Division, Los Alamos National Laboratory, Los Alamos, 
NM 87545}
\author{Joel D. Kress}
\affiliation{Theoretical Division, Los Alamos National Laboratory, Los Alamos, 
NM 87545}
\author{Niels Gr{\o}nbech-Jensen}
\affiliation{Department of Applied Science, University of California, Davis, 
CA 95616}

\date{\today}
\narrowtext
\begin{abstract} 
A new classical potential for methyl-thiol on a Au(111) surface has been 
developed using  density functional theory electronic structure calculations. 
Energy surfaces between methyl-thiol and a gold surface were investigated
in terms of symmetry sites and thiol density. Geometrical 
optimization was employed over all the configurations  while minimum energy
and thiol height were determined. Finally, a new interatomic potential 
has been generated as a function of thiol density, and applications to 
coarse-grained simulations are presented.
\end{abstract}
\maketitle

\section{Introduction}
Understanding the structural features of well-ordered self-assembled monolayers
\cite{dubois92,ulman96,schreiber00,zhang03} requires a detailed potential 
energy surface of alkanethiols adsorbed onto metallic surfaces. 
Thiol headgroups bridge alkane chains with metal surfaces such as gold, 
and play key roles in deciding the properties of surface self-assembly. 
Extensive theoretical 
 studies thereby have been done to explore the thiol interaction with
a gold surface through electronic structure models \cite{Beardmore_98,gronbeck00,hayashi01,yourdshahyan01,yourdshahyan02,morikawa02,gottschalck02,renzi04,masens_05,cometto_05}.
These investigations have revealed that the overall binding energy is 
found as around 40kcal/mol, and the binding energy varies with the relative
sulfur position on the gold lattice. Also, additional experimental 
and computational results indicate  that neighboring thiol configuration 
may affect the thiol-gold binding \cite{schreiber00}. However, the lateral 
energy-corrugation of  the surface by the underlying atomic gold lattice 
has not yet been completely understood.

The purpose of this thiol-gold interaction investigation is two-fold. 
First, the 
self-assembled monolayer of alkanethiol on gold has become a {\it de facto} 
standard for understanding surface self-assembly of micro and 
nanostructures. Profiling the
atomic scale energy surface is fundamental to comprehending the experimental
observations of domain formation and structural defects. Secondly, explicit
description of the energy surface can be formulated as classical potentials, 
which allow large scale molecular dynamics (MD) \cite{jensen03}
or Monte Carlo (MC) simulations.

We have conducted extensive electronic structure calculations of thiol-gold
interactions in order to build a new classical potential, which incorporates
the sulfur position relative to the gold lattice and the many-body effects
from adjacent thiols. This study is an extension of the previous work of
classical potentials for MD and MC simulations \cite{Beardmore_98}.
The previous potential was developed using a dilute thiol model (one thiol
on an isolated gold cluster). 
The results indicated that a fully optimized thiol on 
gold prefers the hollow sites, and the FCC site is more 
stable than HCP. The results have subsequently been verified 
\cite{hick-up_on_misrep}, but later experiments and numerical calculations 
\cite{yourdshahyan01,yourdshahyan02,renzi04} 
have also revealed that Bridge or hybrid sites may provide the 
global minimum. Our present results, which are based on non-zero
thiol density, substantiate these later findings.

The rest of the paper is organized as follows. First we briefly describe
our {\it ab initio} calculations and summarize the results. Then the classical
potential, which has been fitted to the results of the electronic
structure optimizations, will be presented. 
Before concluding the paper, several applications of the new potential within 
MD simulations are exemplified with specific thiol densities on 
a Au(111) surface.

\section{{\it ab initio} calculations} 
Advances in computational capacity and algorithmic efficiency of 
first-principle calculations have greatly improved the quality and 
consistency of numerical simulations of, e.g., thiol on gold. In the
previous study \cite{Beardmore_98}, quantum chemistry techniques with a small
finite size of gold cluster model was employed to produce an energy surface
as a function of thiol position and orientation. However, periodic boundary
conditions (PBC), which imitates a non-zero thiol density, can simulate more 
realistic configurations and is found to efficiently produce reliable results 
\cite{gronbeck00,hayashi01,yourdshahyan02,cometto_05}. We have therefore
employed the band structure simulation code VASP \cite{kresse96c,kresse96} 
to conduct density functional theory (DFT) \cite{hohenberg64,kohn65} 
simulations of various energy surfaces. The Projector Augmented Wave 
(PAW) \cite{blochl94,kresse99} potential was used for electron-ion 
interactions and the Generalized Gradient Approximation (GGA) 
\cite{perdew92} was applied for exchange correlation with a plane wave 
basis. For plane waves, a 400eV cutoff energy was implemented.
Even though plane periodic images are included in the calculation to simulate
dense thiol states, the effect from surface normal periodicity was excluded
by inserting 10{\AA} vacuum layer above the unit-cell.
We used $4\times 4 \times 1$ k-points for the Monkhorst-Pack
grid. With those configurations, {\it ab initio} simulations
were done for various thiol densities and positions.

\begin{figure}
\centerline{\includegraphics[clip, scale=0.5, angle=0]{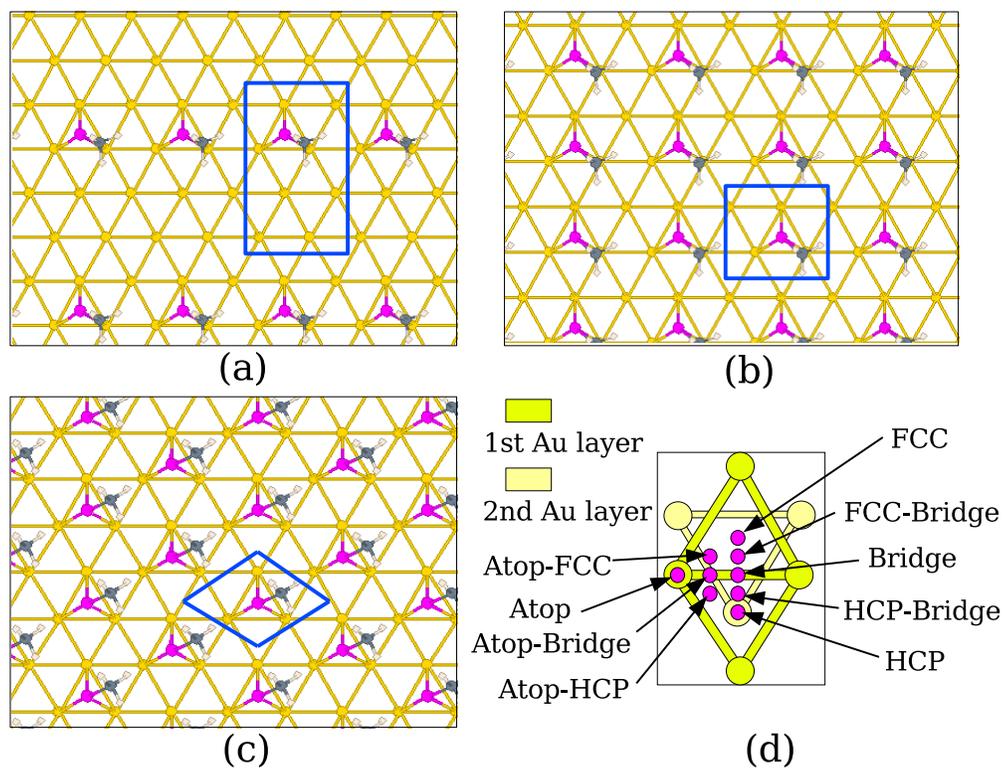}}
\caption{(Color online) Unit-cells with periodic boundary condition: 
(a) $2\!\times\! 4$ cell, (b) $2\!\times\! 2$ cell, 
(c) $ \sqrt 3 \!\times \!\sqrt 3 \rm R 30^o$, (d) nine sampling points for 
landscape variations on a Au(111) surface. }
\label{fig:unitcell}
\end{figure}

With periodic boundary conditions, several surface unit-cells were investigated
in order to mimic thiol density effects. The specific periodicities are:
$2\!\times\!4$, $2\!\times\!2$, and $\sqrt{3}\!\times \!\sqrt{3}\rm R30^o$ 
of Au(111) unit-cells with a single methyl-thiol.  The lattices  are 
depicted in Figure \ref{fig:unitcell}. Figure \ref{fig:unitcell}-(a) shows the 
$2\!\times\!4$ unit-cell with 8 gold atoms per single gold layer.
Figure \ref{fig:unitcell}-(b) shows the $2\!\times\!2$ unit-cell with
gold layers of 4 atoms. The $\sqrt{3}\!\times \!\sqrt{3}\rm R30^o$ unit-cell, 
shown in Figure \ref{fig:unitcell}-(c), represents the maximum
thiol packing density (full coverage) of a Au(111) surface \cite{ulman96} (Note
that the precise location of the unit-cells are not accurately represented
in Figure \ref{fig:unitcell}). 

In order to evaluate the energy surface on symmetric sites, nine sampling 
points were selected, including Atop, Bridge, FCC, HCP, and hybrid points
in between. These points are illustrated in Figure \ref{fig:unitcell}-(d). 

The methyl-thiol binding energy to the Au(111) surface is determined by 
comparing the energies of relaxed models of Au(111) with and without the
methyl-thiol:
\begin{equation}
E_{\rm bind} = E_{\rm Au(111)+SCH_3} - E_{\rm Au(111)} - E_{\rm SCH_3}
\end{equation}

\subsection{Effect of Au(111) thickness}
\begin{figure}
\centerline{\includegraphics[clip, scale=0.8, angle=0]{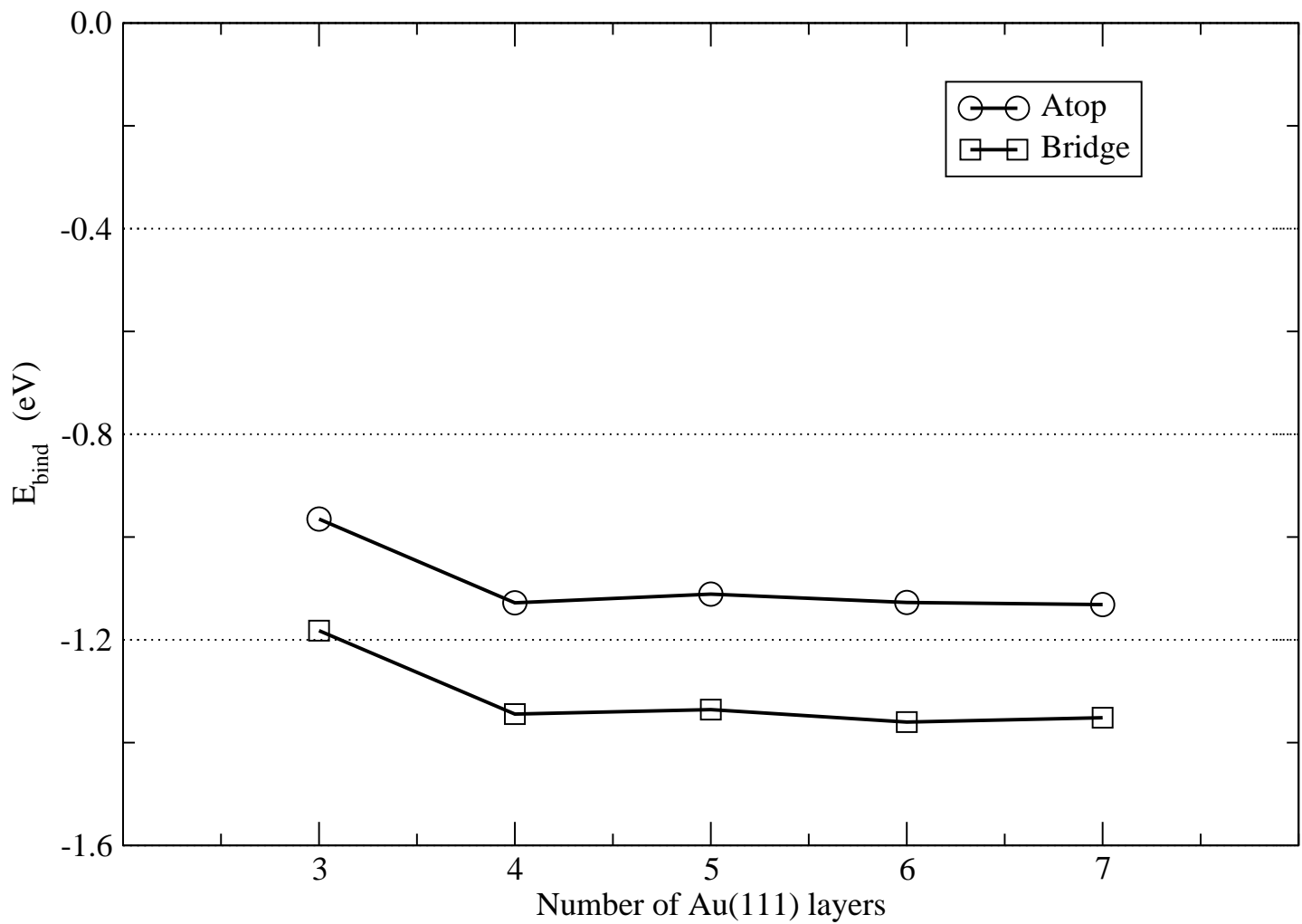}}
\caption{The effect of gold layer thickness on the binding energy of 
methyl-thiol on a Au(111) surface. For more than four layers, 
binding energies are saturated. }
\label{fig:au_layer}
\end{figure}

\begin{figure}
\centerline{\includegraphics[clip, scale=0.5, angle=0]{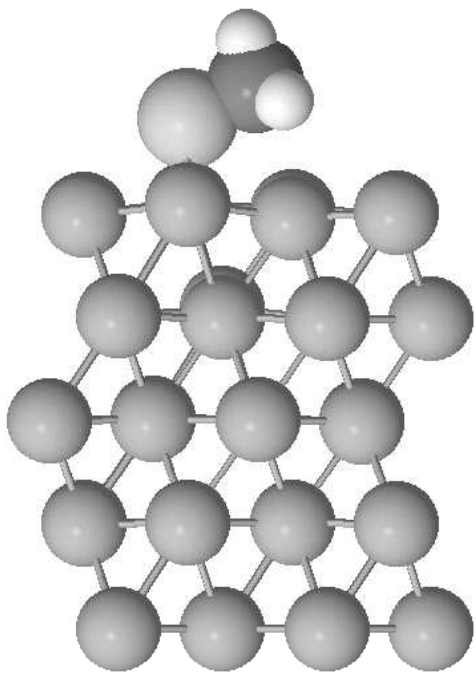}}
\caption{Side view of methyl-thiol on Au(111) of a five-layer slab.}
\label{fig:5layer}
\end{figure}

It is known that a small number of atomic layers of gold will influence the
calculated binding energy \cite{yourdshahyan01}. We therefore determine an
acceptable number of gold layers as follows:
Atop and Bridge sites are found as global maximum and
minimum \cite{yourdshahyan01,yourdshahyan02}, 
and can therefore serve as standard points for a gold layer effect study.
Atop and Bridge sites on a $\sqrt{3}\!\times \!\sqrt{3}\rm R30^o$ unit-cell
were configured with various numbers of Au(111) layers (with lattice 
constant=4.08 {\AA}), 
and tested by the {\it ab initio} calculations described above. 
The bottom layer was fixed and upper layers were relaxed along the
normal direction. Simulation results are provided in
Figure \ref{fig:au_layer} where the effect of the number of gold layers 
is illustrated. 
More than four layers, binding energies for both points saturate.
Also, the energy difference between the two points is
maintained when more than four gold lattice layers are simulated. 
We conclude that at least four layers are required to adequately calculate
the binding energy. We employed five Au(111) layers for the binding energy
calculations of above mentioned nine sampling points, as illustrated in 
Figure \ref{fig:5layer}.

\subsection{{\it ab initio} calculation results}
\begin{figure}
\centerline{\includegraphics[clip, scale=0.8, angle=0]{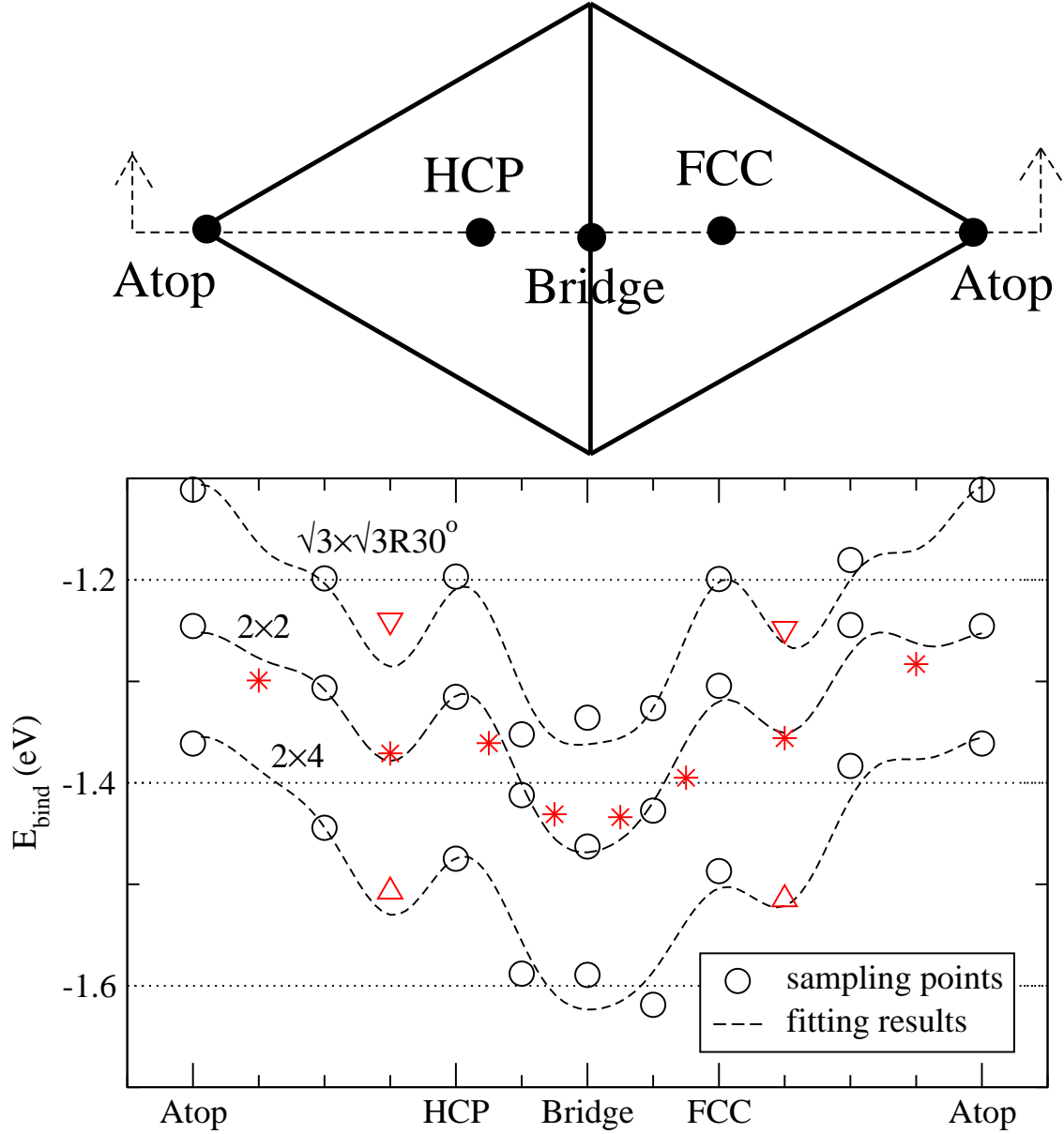}}
\caption{(Color online) {\it ab-initio} results of sampling points and the
fitting results. Down triangles are the results of fine analysis of  
$ \sqrt 3 \!\times \!\sqrt 3 \rm R 30^o$ unit-cell, while 
asterisks(*) for $2\!\times\!2$ and up triangles for 
$2\!\times\!4$ unit-cell. These test points are not included in the fitting.}
\label{fig:energy}
\end{figure}

All nine points [Figure 1-(d)] on the Au(111) surface were probed
with different thiol densities, and the results are summarized in Table 
\ref{table:results}. Also selected results are provided in 
Figure \ref{fig:energy} along the Atop-Bridge-Atop path. 
Even though thiol density affects the magnitude
of binding energies, lateral corrugations of the energy were found to be
similar for all three densities. The Atop site was the least 
stable and the FCC sites were slightly
more stable than HCP although the difference is negligible. 
Finally, Bridge and hybrid sites of Bridge-FCC and Bridge-HCP
present the strongest binding energy, in agreement with other recent studies
\cite{hayashi01,yourdshahyan02,gottschalck02,cometto_05}.
Overall binding energy increases as thiol density decreases, while 
lateral corrugation seems largely unaffected. Even though global minimum
points move from HCP-Bridge site to FCC-Bridge site, but the difference is not
distinct. In other words, global minimum and maximum sites are not 
significantly changed by the variations of thiol density, whereas 
local thiol density determines the magnitude of the 
binding energy. Thus, thiol-position and local thiol-density
can be decoupled in a classical potential function such that a many-body
component to the energy surface provides mutual thiol repulsion.

In addition to the binding energies, optimized vertical distances 
between thiol and Au(111) surface were investigated. Similarly to the 
binding energy results, an Atop positioning of thiol showed the largest 
distance, while optimized thiol  distances to FCC and HCP sites are small.
An interesting point is that these results are quite similar regardless of
thiol density. Consequently, our fitting of the thiol-Au(111) distance
surface was done without a thiol density term. 

\section{Generation of energy surfaces of classical potentials}
Based on the previous work \cite{Beardmore_98}, we have interpolated and
fitted the electronic structure results with harmonic functions to 
compose an egg-box type surface, which can reproduce three-fold symmetry.
With following characteristic vectors and harmonic sums, 
all the 3-fold symmetric sites of the Au(111) surface
are periodically represented:
\begin{eqnarray}
 {\bf k}_1 = k_0 \left( {\sqrt 3 \over 2}, {1\over 2} \right), ~~{\bf k}_2 = k_0 \left(
-{\sqrt 3 \over 2}, {1\over 2} \right),~~{\bf k}_3 = k_0 \left( 0,-1 \right), 
\label{eqn:k}
\end{eqnarray}
where $k_0 = 4\pi /  {\sqrt 3} a$, and $a$ is the inter-atomic distance for
gold (=2.884{\AA}).
Using basic and high order harmonic functions, energy surfaces were determined.
The fitting functions for the binding energies are shown below. 

\begin{eqnarray}
E_{\rm bind} &=&a_0 + a_1\sum_{i=1}^3\cos ({\bf k}_i \cdot{\bf x})+ 
a_2\sum_{i=1}^3\sin ({\bf k}_i \cdot{\bf x}) \nonumber 
 + a_3\sum_{i=1}^3\cos (2{\bf k}_i \cdot{\bf x}) \\
& & + a_4\sum_{i=1}^3\sin (2{\bf k}_i \cdot{\bf x}) \label{eqn:surfe_fit} 
 + a_5\sum_{i=1}^3\cos (3{\bf k}_i \cdot{\bf x}) 
. 
\label{eqn:generic}
\end{eqnarray}

For high-order harmonics, we included up to third harmonics but the {\bf sin} 
function of third order was excluded due to negligible contributions.
All fitting coefficients are summarized in 
Table \ref{table:fitting_energy} for each thiol density. 
They are interpolated using a cubic spline, and surface coefficients for
an intermediate density are given as
\begin{equation}
a_i = s_3 \cdot\bar\rho^3 + s_2 \cdot\bar\rho^2 +s_1 \cdot\bar\rho +  s_0 .
\label{eqn:cubic}
\end{equation}

Cubic spline coefficients $s_i$ are provided in Table \ref{table:cubic_e}, 
where the thiol density is normalized to that of the $2\!\times\!2$ unit-cell. 

With the above equations, the binding energy is addressed in terms of 
thiol density and position. The comparison to the sampling points are 
illustrated in Figure \ref{fig:energy}. Rms error of 27 sampling points was
calculated as 0.018 eV. The fitted energy surface is seen to represent the 
sampling point data quite well with the selected level of harmonic
functions. However, the fitted function has variations that are not
directly given by the sample points. In order to investigate if this
variation is an artifact of the fitting function, we have chosen to
compare the fitting result to twelve additional energies derived at
intermediate thiol surface locations: eight additional for one of the thiol
densities (2x2 unit-cell) and two additional for each of the other two
densities.  These are represented with asterisks (*) and triangles. 
It is evident from the comparison in Figure 4 that the fitted
energy curve represents also the independent validation points very
well, and so, we submit that the presented fitting is representative
of the surface generated from ab initio methods.

\begin{figure}
\centerline{\includegraphics[clip, scale=0.8, angle=0]{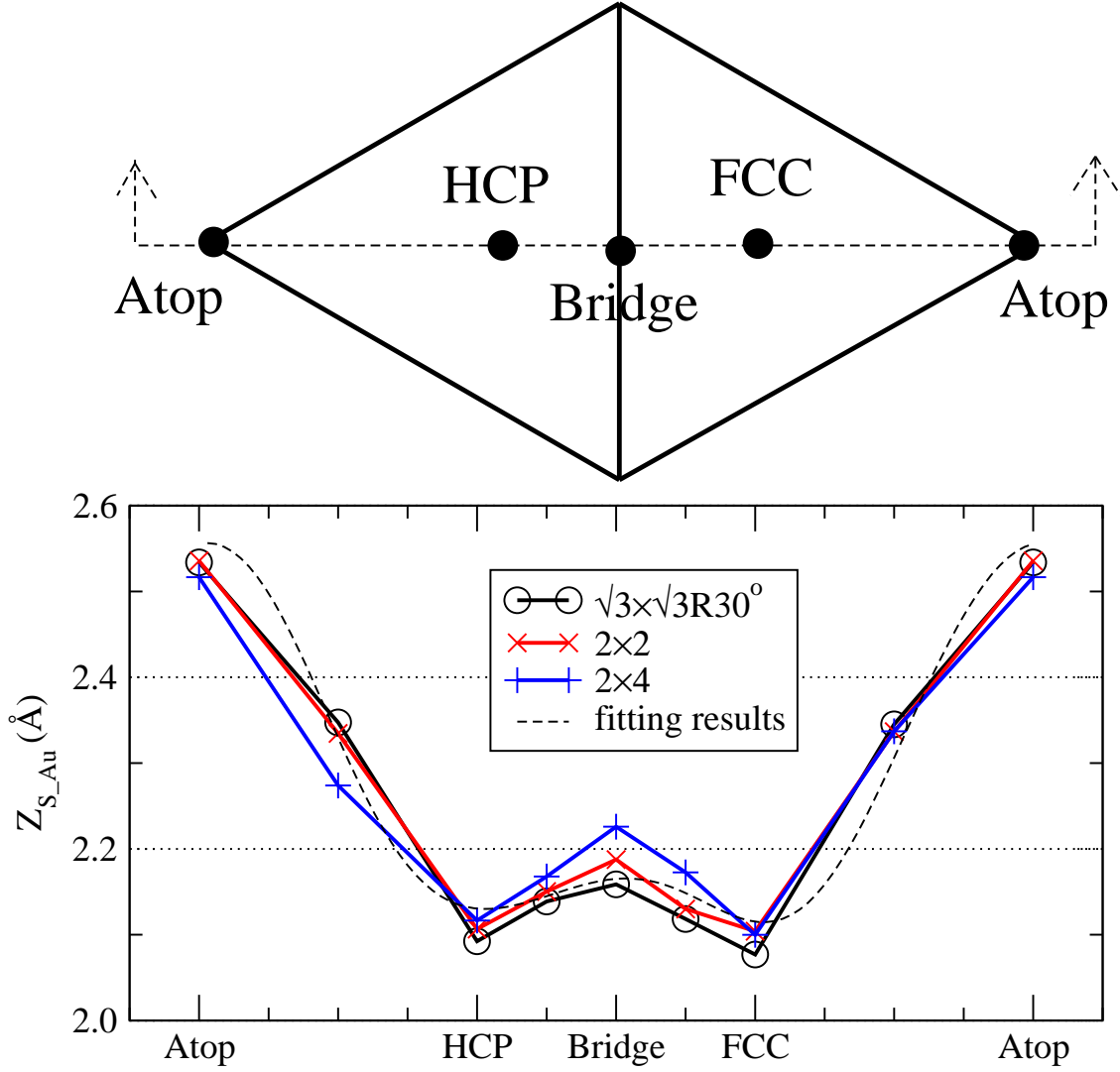}}
\caption{(Color online) The thiol-Au(111) distance along symmetric sites. 
The effect of thiol density change is not distinct in the thiol-Au(111) 
distance, and a single fitting surface is employed. }
\label{fig:surface}
\end{figure}

The fitting method was also used for the optimized thiol-Au(111) distance 
surface. The fitting functions are shown below and a comparison to the 
{\it ab initio} data is plotted in Figure \ref{fig:surface}.

\begin{eqnarray} 
Z_{\rm S\_Au} &=&b_0 + b_1\sum_{i=1}^3\cos ({\bf k}_i \cdot{\bf x})+ 
b_2\sum_{i=1}^3\sin ({\bf k}_i \cdot{\bf x}) \nonumber \\
& & + b_3\sum_{i=1}^3\cos (2{\bf k}_i \cdot{\bf x})+ 
b_4\sum_{i=1}^3\sin (2{\bf k}_i \cdot{\bf x}) \label{eqn:surf_z} 
\label{eqn:z}
\end{eqnarray}

\begin{figure}
\centerline{\includegraphics[clip, scale=1.0, angle=0]{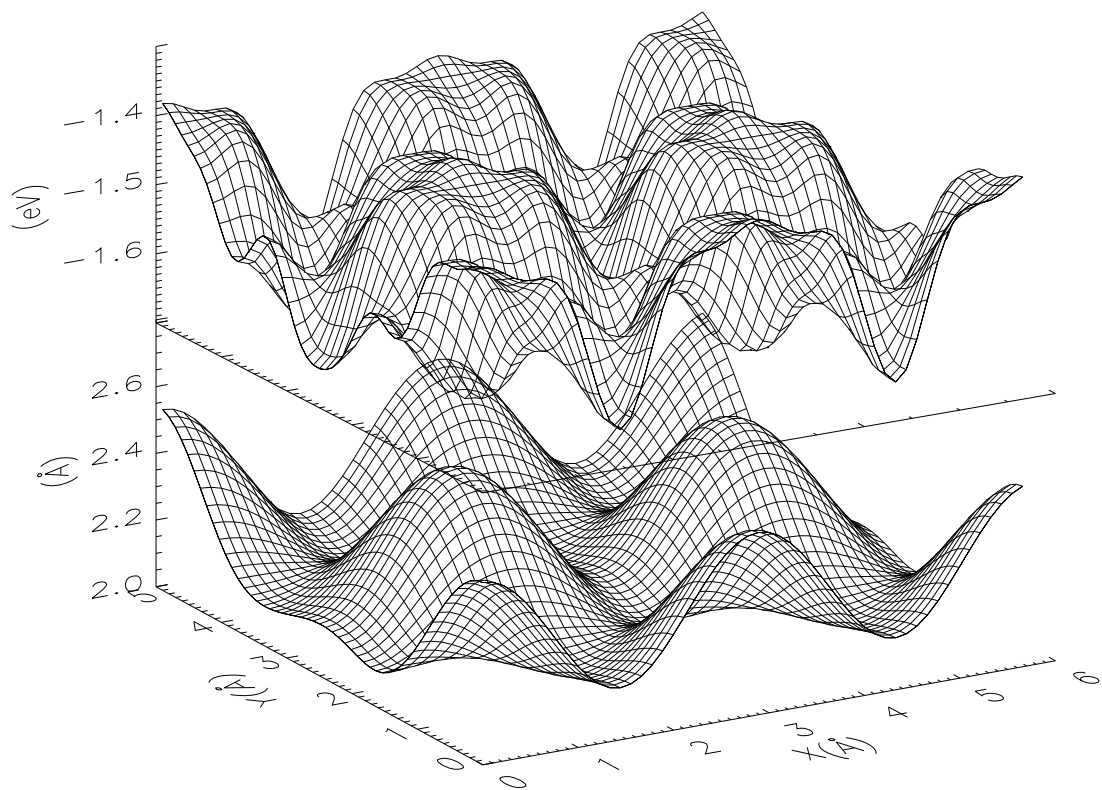}}
\caption{ (Upper) Energy surface for $2 \!\times \!4$ cell. For the other 
thiol density models, the shapes of energy surfaces are consistent with 
some changes of magnitude. (Lower) Thiol-Au(111) distance surface.  }
\label{fig:sample}
\end{figure}

Unlike the binding energy, this fitting is not affected by thiol density and
it is therefore shown in terms of thiol position only. Fitting coefficients 
of the distance
surface are summarized in Table \ref{table:s_au}. From the fitting results,
a sample energy surface of  a $2\!\times\!4$ unit-cell and
the corresponding thiol-Au(111) distance surface is illustrated in 
Figure \ref{fig:sample}. 
Atop sites are located (in \AA) at (0,0), (2.884,0), and (1.442,2.498) while
HCP is at (1.442,0.833). All other symmetric sites can be mapped along three
fold symmetry. Atop sites are seen to be energetic maxima  whereas thiols near
Bridge sites show the strongest binding to the surface. 
Other thiol densities have similar energy surfaces.
On the distance surface, peaks are found around Atop, but the Bridge site 
is slightly higher than the hollow sites. FCC and HCP show similar heights. 
We notice, however, that the variation in height observed in Figure 6, is less
than {1\AA}, indicating that the height variation is not an important 
component for describing surface self-assemblies.

In essence, the fitting coefficients of energy surfaces are determined 
by a local thiol density, and we have to collect the effects of neighboring 
thiols. The embedded atom method (EAM) \cite{daw84} employs the local 
density to incorporate many-body effects into classical MD potentials.
For that purpose, we determine the local thiol density $\rho_i$ as
\begin{equation}
 \rho_i = \sum_j \rho ( r_{ij}),
\label{eqn:rhosum}
\end{equation}
where $r_{ij}$ is the distance between the given thiol {\it i} and all 
adjacent thiols {\it j}. Each neighbor contributes to the density $\rho_i$ of 
the {\it i}th thiol through the function $\rho$. 
Using the following function, the local thiol density can be defined as
\begin{eqnarray}
 \rho(r_{ij}) & = & \left\{\begin{array}{clr}  \big\{ 1 - \cos \left( {2\pi 
(r_{ij} - 5 a) \over 5 a} \right) \big\}/2 ~~~ & {\rm if} & r_{ij} < 5a \\
 0 &  {\rm if} & r_{ij} \ge 5a
\end{array} \right. ,
\label{eqn:density}
\end{eqnarray}
where $a$ is the  inter-atomic distance for gold (=2.884{\AA}). The formulation
of local density reproduces the corresponding densities  with our chosen 
unit-cells: $2\!\times\!4$, $2\!\times\!2$, and 
$\sqrt{3}\!\times \!\sqrt{3}\rm R30^o$. We therefore adopt the 
form (\ref{eqn:density}) to include a thiol density into the calculation 
of the binding energy ($E_{\rm bind}$).

We finally combine the local thiol density and position into a Morse potential:
\begin{eqnarray}
V(\rho_i, {\bf x}_i, z_i ) & = & E_{\rm bind} (\rho_i , {\bf x}_i) 
\cdot \Big\{ \exp[ - 2 \beta_e (z_i - Z_{\rm S\_Au}({\bf x}_i ))] 
\nonumber  \Big. \\
 & & - 2  \exp [-\beta_e(z_i - Z_{\rm S\_Au}({\bf x}_i ))] \Big. \Big\}. 
\label{eqn:morse}
\end{eqnarray}

Here,
$z_i$ is the normal distance between the thiol and the Au(111) surface at the 
position ${\bf x}_i$,  and $\beta_e$ is the curvature of the potential. To 
determine the potential curvature, perturbations to the thiol-Au distances were
given on the optimized states. Energy differences and the gradient
were calculated along the distance variations. It was found empirically 
that the gradient is proportional to   $Z_{\rm S\_Au}$ as
\begin{equation}
\beta_e = c_1 Z_{\rm S\_Au}.
\label{eqn:beta}
\end{equation}

The constant $c_1$ is a function of the local thiol density and 
calculated with a cubic spline similarly to Eq.(\ref{eqn:cubic}). 
All parameters are summarized in Table. \ref{table:cubic_beta}.

\section{Application to classical molecular dynamics simulations}
In order to demonstrate the practical applications of the developed potential, 
we have simulated methyl-thiol headgroup distributions on Au(111) for several
surface coverages. Classical MD  has been implemented with a united atom 
force field \cite{hautman89} for the chain-chain interactions, and 
RATTLE \cite{andersen83} for  constraints of  the carbon-carbon bond lengths 
and angles. The torsional potential for dihedral angles were adopted 
from \cite{hautman89}, and the thiol headgroup potential was incorporated 
using Eq.(\ref{eqn:generic}).

Decanethiols are distributed along the simulation cell of 
{ 86.52\AA $\times$ 79.92 \AA} and periodic boundary conditions are imposed. 
Coverages are 1/2, 2/3, and 1 of optimal coverage with 160, 210, and 320 
decanethiols, respectively. A stochastic thermostat \cite{hunenberger05} was 
set at 300K and the system was relaxed for 100ps. 

\begin{figure}
\centerline{
\includegraphics[clip, scale=0.5, angle=0]{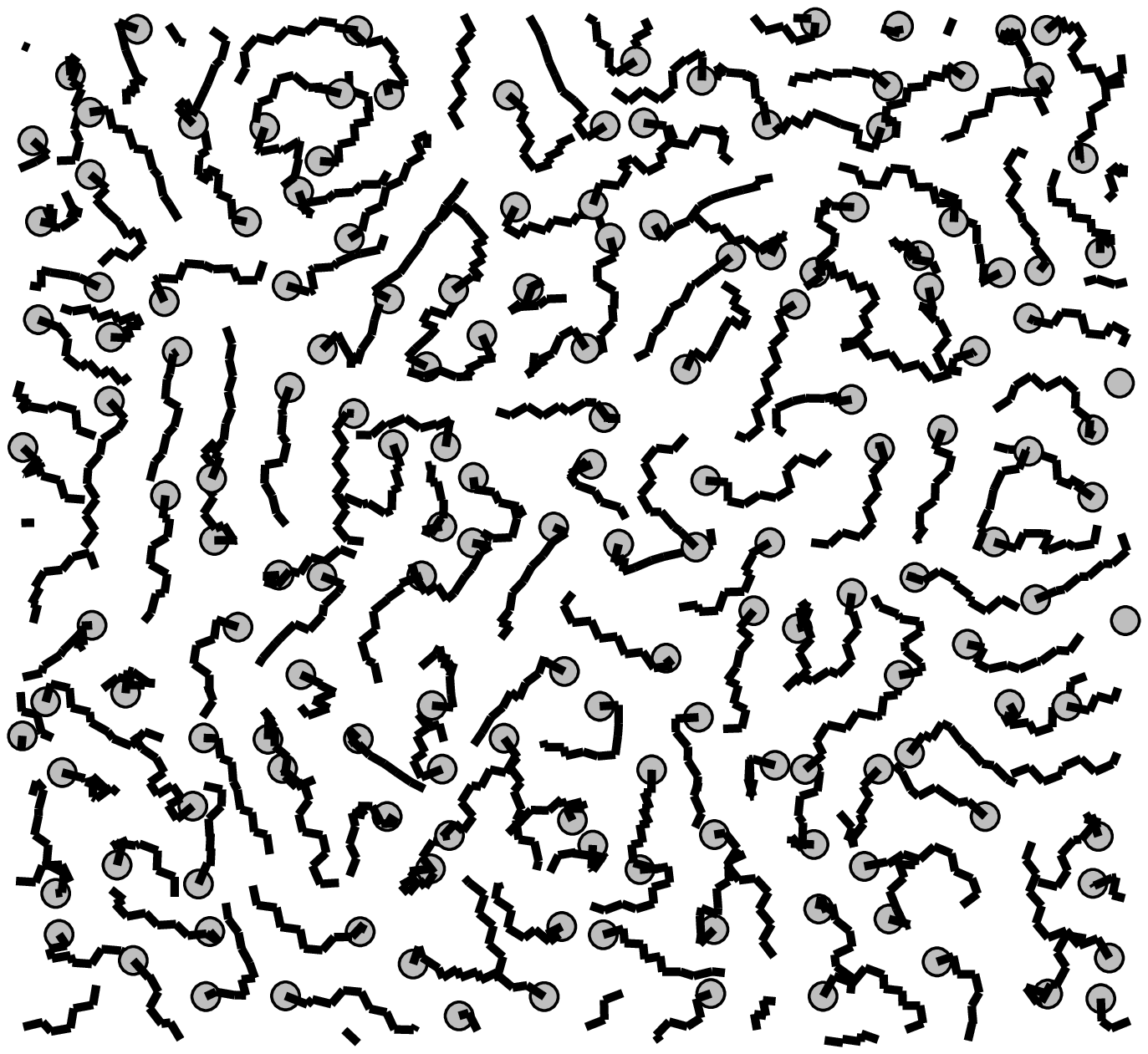}
\includegraphics[clip, scale=0.5, angle=0]{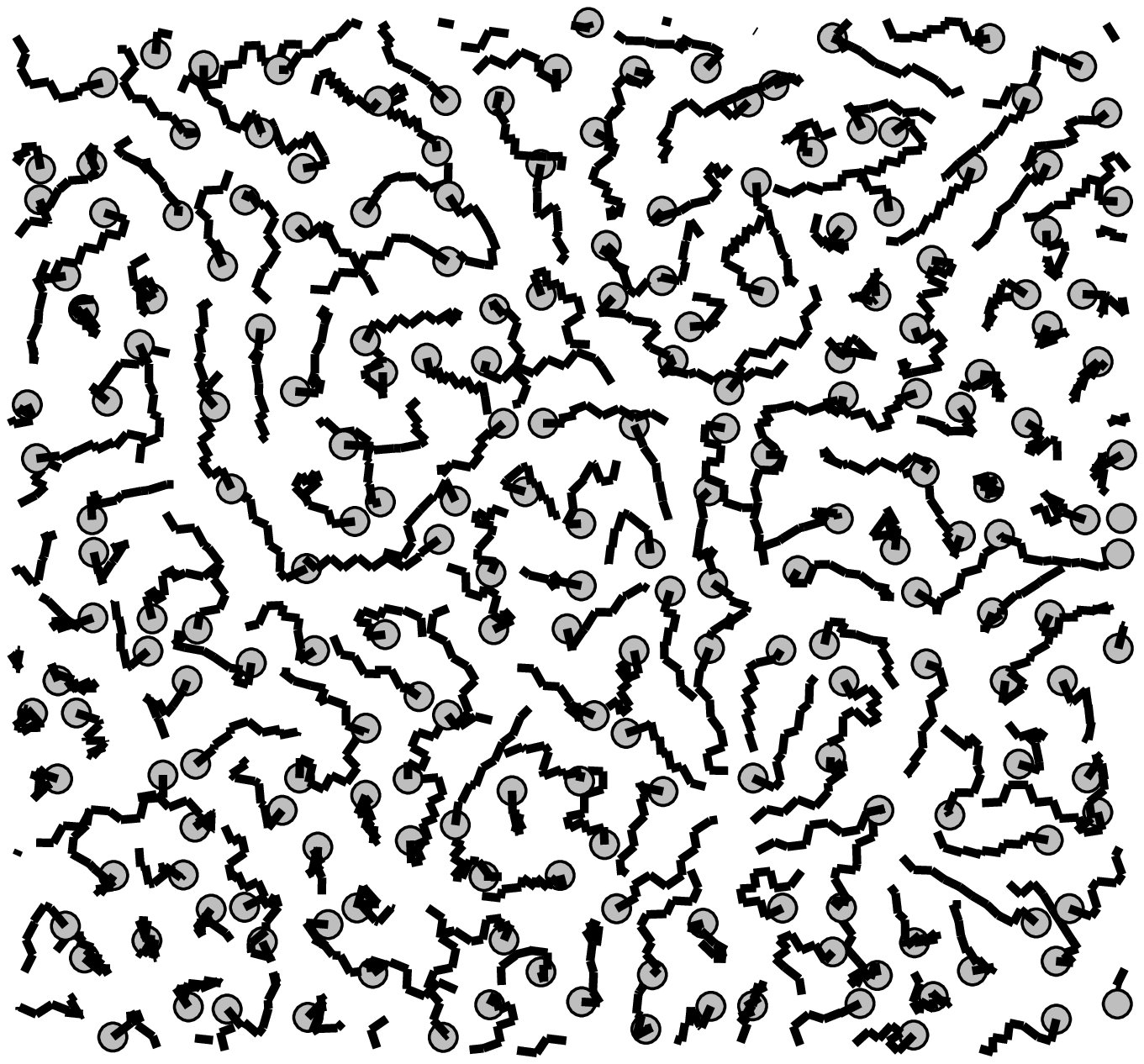}
\includegraphics[clip, scale=0.5, angle=0]{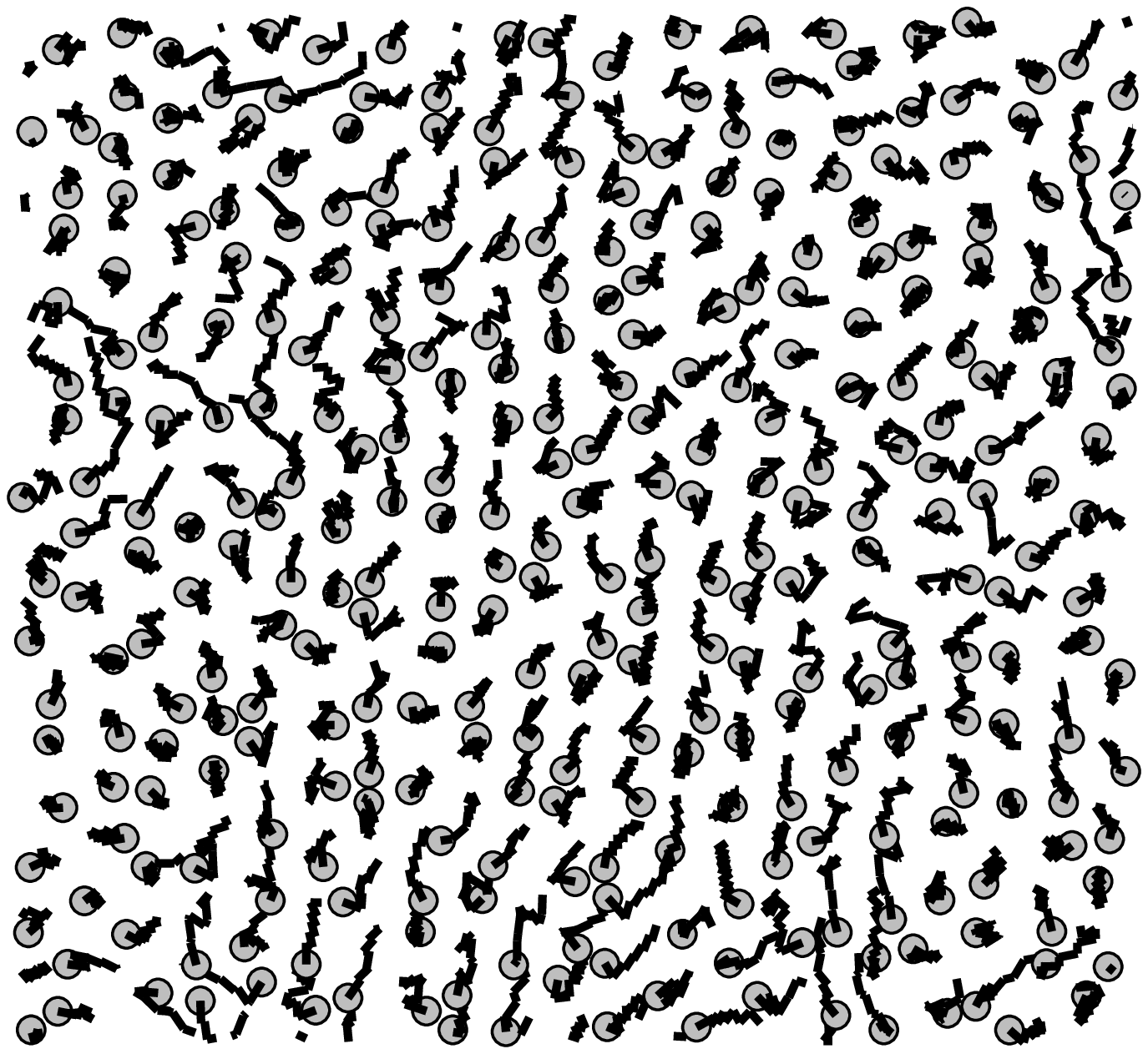}
}
\caption{Top view of alkane chains (black line) and thiol headgroups 
(grey circle) on gold surface from MD simulations 
(left: 1/2 coverage, middle: 2/3 coverage, right: full coverage).}
\label{fig:md}
\end{figure}

\begin{figure}
\centerline{\includegraphics[clip, scale=0.8, angle=0]{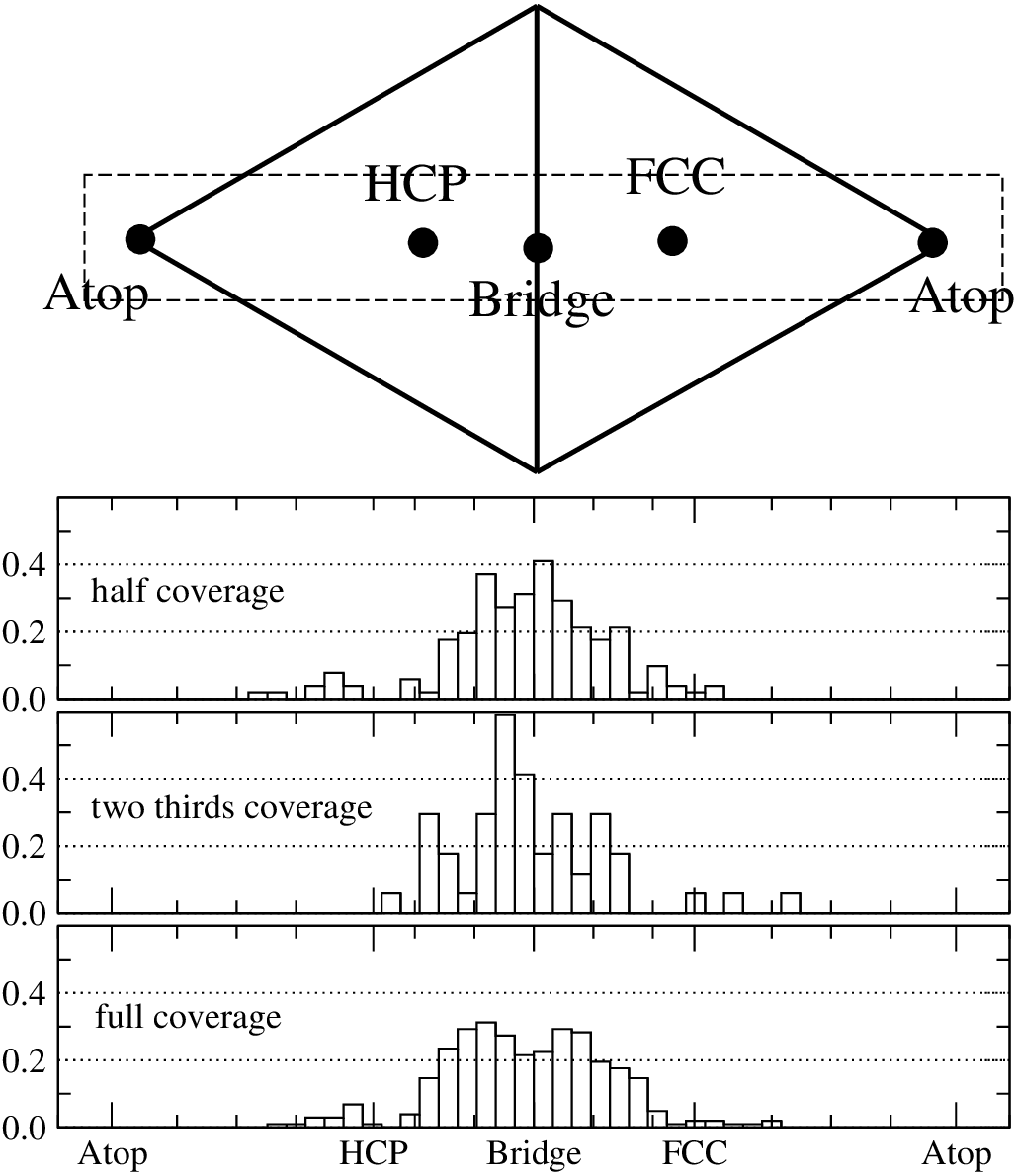}}
\caption{Distribution of headgroups along symmetric sites from MD simulations.
They are normalized distributions and consistently headgroups are found around 
Bridge site with high probabilities.}
\label{fig:head2}
\end{figure}

Figure \ref{fig:md} shows top views of the MD results using the developed 
headgroup potentials. At low density the alkane chains lie between 
neighboring chains and the headgroups show irregular patterns. As coverage 
increases, alkane chains interact with each other and lift from the surface. 
At high density the headgroups order hexagonally, and alkane chains tilt 
toward NN (Nearest Neighbor) or NNN (Next Nearest Neighbor). In order to 
investigate headgroup distributions along the characteristic line 
(Atop-HCP-Bridge-FCC-Atop), their locations at each coverage were examined 
and normalized histograms are presented in Figure \ref{fig:head2}. For all 
coverages, most of the headgroups are found between HCP-Bridge and FCC-Bridge 
hybrid sites. This result is consistent with the energetics of the electronic 
structure calculations.

\section{Conclusion and discussion}
In summary, electronic structure calculations have been conducted to 
characterize the details of interactions between methyl-thiol and Au(111) 
surfaces. The energy surface has been studied as a function of thiol 
location and density. The results are consistent
with other recent work, and a new classical potential has been developed
from the results. The potential is completely parametrized, and its 
application to MD is demonstrated.
Large scale molecular
dynamics simulations are under way to explore the domain formation and
structure composition of surface self-assembly using the surface potential 
developed in this paper.

\subsection{Acknowledgments}
This work was carried out under the auspices of the National Nuclear Security
Administration of the
U.S. Department of Energy at Los Alamos National Laboratory under
Contract No. DE-AC52-06NA25396 and the
Cooperative Agreement on Research and Education (CARE) for UCDRD funding.
Partial support was provided by the National Science Foundation 
Biophotonics Science and Technology Center (University of California at
Davis).


\begin{thebibliography}{28}
\bibitem[{hic()}]{hick-up_on_misrep}
\bibinfo{note}{The favored gold(111) fcc hollow site positioning of single
  thiols has been reconfirmed by more recent extensive {DFT} calculations in,
  for example, ref\cite{yourdshahyan01,yourdshahyan02}. Unfortunately, those
  publications have erroneously quoted ref \cite{Beardmore_98} as suggesting
  the hcp hollow site is the energetically preferred location of thiols}.
\end{thebibliography}

\clearpage
\begin{table}[t]
\caption{{\it ab initio} results  of binding energy ($E_{\rm bind}$) 
in terms of symmetric sites for each  unit-cell.}
\begin{center}
\begin{tabular}{c|c|c|c}
\hline
unit-cell & $~~~~2\! \times\! 4~~~~$ & $~~~~2\! \times \!2~~~~$ & $\sqrt3 \!\times \!\sqrt3 \rm R 30^o$ \\
\hline \hline
ATOP        & -1.361 & -1.245 & -1.111\\
Bridge      & -1.589 & -1.463 & -1.336\\
FCC         & -1.487 & -1.304 & -1.199\\
HCP         & -1.475 & -1.315 & -1.197\\
Atop-Bridge & -1.402 & -1.290 & -1.158\\
Atop-HCP    & -1.444 & -1.306 & -1.199\\
Atop-FCC    & -1.383 & -1.245 & -1.181 \\
FCC-Bridge  & -1.619 & -1.427 & -1.326 \\
HCP-Bridge  & -1.586 & -1.412 & -1.352 \\

\hline
\end{tabular}
\end{center}
\label{table:results}
\end{table}

\clearpage
\begin{table}[t]
\caption{Egg-box fitting parameters of binding energy ($E_{\rm bind}$) for each  unit-cell.}
\begin{center}
\begin{tabular}{c|r|r|r}
\hline
unit-cell & $2\!\times\! 4~~~~~$ & $2 \!\times\! 2~~~~~~ $ & $\sqrt3 \!\times \!\sqrt3 \rm R 30^o$ \\
\hline \hline
{  $a_0$} &   $-1.481\times 10^{-0}$ &   $-1.335\times 10^{-0}$ & 
 $ -1.233\times 10^{-0}$ \\
{  $a_1$} &  $5.451\times 10^{-2} $ &   $4.009\times 10^{-2} $ &
$4.285\times 10^{-2} $ \\
{  $a_2$} & $3.218\times 10^{-3}$ & $6.910\times 10^{-3}$ &
 $4.236\times 10^{-3} $ \\
{  $a_3$} &  $-2.497\times 10^{-2} $ &  $ -2.607\times 10^{-2} $ &
  $ -2.134\times 10^{-2} $ \\
{  $a_4$} & $9.041\times 10^{-3}$  &  $ 8.068\times 10^{-3}$  & 
 $2.897\times 10^{-3}$  \\ 
{  $a_5$} & $1.266\times 10^{-2}$  &  $1.375\times 10^{-2}$ &
  $2.093\times 10^{-2} $ \\
\hline
\end{tabular}
\end{center}
\label{table:fitting_energy}
\end{table}

\begin{table}[b]
\caption{Cubic spline coefficients for egg-box fitting parameters of binding 
energy (in eV). $ \bar \rho $ = normalized thiol density. }
\begin{center}

\begin{tabular}{c|r|r|r|r}
\hline
& \multicolumn{4}{c}{  $0.5 < \bar \rho < 1.0$} \\
\hline
\hline
 & $s_3~~~~~~~~~$ & $s_2~~~~~~~~~$& $s_1~~~~~~~~~$& $s_0~~~~~~~~~$ \\
\hline
$a_0 $ &  $  1.760\times 10^{-2} $ &  $  -2.641\times 10^{-2} $ &  
  $3.005\times 10^{-1}$ &   $ -1.627\times 10^{-0}$ \\
$a_1 $ &   $ 4.458\times 10^{-2} $ &   $ -6.686\times 10^{-2} $ & 
   $-6.546\times 10^{-3}$ &   $ 6.892\times 10^{-2}$ \\
$a_2 $ &    $-1.851\times 10^{-2} $ &  $  2.776\times 10^{-2} $ &  
  $-1.871\times 10^{-3}$ &    $-4.737\times 10^{-4} $ \\
$a_3 $ &  $  1.968\times 10^{-2} $ &    $-2.952\times 10^{-2} $ & 
  $ 7.647\times 10^{-3}$ &    $-2.387\times 10^{-2}$ \\
$a_4 $ &  $  -1.631\times 10^{-2} $ &    $2.447\times 10^{-2} $ & 
 $  -1.010\times 10^{-2} $ &   $ 1.001\times 10^{-2}$ \\
$a_5 $ &  $  2.326\times 10^{-2} $ &    $-3.488\times 10^{-2} $ & 
$   1.382\times 10^{-2} $ &   $ 1.156\times 10^{-2}$  \\
\hline
& \multicolumn{4}{c}{ $1.0 < \bar \rho < 1.33$} \\
\hline
\hline
& $s_3~~~~~~~~~$ & $s_2~~~~~~~~~$& $s_1~~~~~~~~~$& $s_0~~~~~~~~~$ \\
\hline
$a_0 $ &
$-2.644\times 10^{-2} $ &   $  1.057\times 10^{-1} $ &   
$  1.684\times 10^{-1}$ &   $  -1.583\times 10^{-0} $ \\
$a_1 $ &
$-6.696\times 10^{-2} $ &   $  2.677\times 10^{-1}$ &   
 $ -3.411\times 10^{-1}$ &  $   1.805\times 10^{-1}$ \\
$a_2 $ &
$2.780\times 10^{-2} $ &    $ -1.112\times 10^{-1} $ &   
 $ 1.371\times 10^{-1}$ &    $ -4.678\times 10^{-2} $ \\
$a_3 $ &
$-2.956\times 10^{-2} $ &  $   1.182\times 10^{-1}$ &  
 $  -1.401\times 10^{-1}$ &   $  2.537\times 10^{-2} $ \\
$a_4 $ &
$2.450\times 10^{-2} $ &    $ -9.798\times 10^{-2} $ &  
 $  1.123\times 10^{-1}$ &   $  -3.080\times 10^{-2} $ \\
$a_5 $ &
$-3.493\times 10^{-2} $  &     $   1.397\times 10^{-1}$  &   
  $  -1.607\times 10^{-1}$  &    $  6.975\times 10^{-2} $ \\

\hline
\end{tabular}
\end{center}
\hfill
\label{table:cubic_e}
\end{table}

\clearpage
\begin{table}[t]
\caption{Egg-box fitting parameters of the thiol-Au distance surface ($Z_{\rm S\_Au}$) for each unit-cell (unit:{\AA}). }
\begin{center}

\begin{tabular}{c|r|r|r|r}
\hline

$b_0$ & $b_1~~~~~~$ & $b_2~~~~~~$ & $b_3~~~~~~$  & $b_4~~~~~~$ \\
\hline
$  2.267\times 10^{-0} $ &    $9.771\times 10^{-2} $ &   
$5.552\times 10^{-4}$ &   $-1.262\times 10^{-3}$ & $3.516\times 10^{-3}$ \\
\hline

\end{tabular}
\end{center}
\label{table:s_au}
\end{table}

\begin{table}[b]
\caption{Cubic spline coefficients for the curvature of thiol potential. 
$ \bar \rho $ = normalized thiol density. }
\begin{center}

\begin{tabular}{c|r|r|r|r}
\hline
& \multicolumn{4}{c}{ $ 0.5 < \bar \rho < 1.0 $}  \\
\hline
 & $s_3~~~~~$ & $s_2~~~~~$& $s_1~~~~~$& $s_0~~~~~$ \\
\hline
$c_1$ &   $-9.660\times 10^{-3}$ &    $1.449\times 10^{-2} $ & 
   $3.545\times 10^{-2} $ &    $4.536\times 10^{-2} $ \\

\hline
\hline
& \multicolumn{4}{c}{ $ 1.0 < \bar \rho < 1.33 $} \\
\hline
 & $s_3~~~~~$ & $s_2~~~~~$& $s_1~~~~~$& $s_0~~~~~$\\
\hline
$c_1$ &    $1.451\times 10^{-2} $ &    $-5.802\times 10^{-2} $ &    
$1.080\times 10^{-1}$ &    $2.119\times 10^{-2} $ \\
\hline
\end{tabular}
\end{center}
\label{table:cubic_beta}
\end{table}

\clearpage

\clearpage

\clearpage

\clearpage

\clearpage

\clearpage

\clearpage

\clearpage

\end{document}